\documentclass[sigplan]{acmart}
\usepackage[utf8]{inputenc} %
\usepackage[T1]{fontenc}    %
\usepackage{hyperref}       %
\usepackage{url}            %
\usepackage{booktabs}       %
\usepackage{nicefrac}       %
\usepackage{microtype}      %
\usepackage{xcolor}         %

\usepackage{mathtools}
\usepackage{siunitx}
\usepackage[abbreviations]{foreign}
\usepackage{multirow}
\usepackage{color, colortbl}

\usepackage{algorithm}
\usepackage[noend]{algpseudocode}

\usepackage{graphicx}
\usepackage{subcaption}

\usepackage{ifthen}
\newboolean{showcomments}
\setboolean{showcomments}{true}
\ifthenelse{\boolean{showcomments}}
{ \newcommand{\mynote}[3]{
		\fbox{\bfseries\sffamily\scriptsize#1}
		{\small$\blacktriangleright$\textsf{\emph{\color{#3}{#2}}}$\blacktriangleleft$}}
	\newcommand{\zzz}[1]{{\setlength{\fboxsep}{2pt}\fcolorbox{black}{yellow}{\textsf{\emph{#1}}}}\xspace}}
{ \newcommand{\mynote}[3]{}
	\newcommand{\zzz}[1]{}}

\usepackage{acronym}
\acrodef{DL}{decentralized learning}
\acrodef{GL}{gossip learning}
\acrodef{EL}{epidemic learning}
\acrodef{ML}{machine learning}
\acrodef{FL}{federated learning}
\acrodef{D-PSGD}{decentralized parallel stochastic gradient descent}
\acrodef{SGD}{stochastic gradient descent}

\newcommand{\sys}{\textsc{Plexus}\xspace}
\newcommand{\FL}{\ac{FL}\xspace}
\newcommand{\DL}{\ac{DL}\xspace}

\newcommand{\femnist}{FEMNIST\xspace}
\newcommand{\celeba}{CelebA\xspace}

\newcommand{\cifar}{CIFAR-10\xspace}

\newcommand{\leaf}{LEAF\xspace}

\newcommand{\dpsgd}{{\xspace}\ac{D-PSGD}\xspace}

\algnewcommand{\LeftComment}[1]{\State {\color{violet} \slash* #1 *\slash}}
\algrenewcommand{\algorithmiccomment}[1]{{\color{violet}\hfill$\triangleright$ #1}}

\algblockdefx[Request]{Request}{EndRequest}{\textbf{request}~}{\textbf{end}}
\makeatletter
\ifthenelse{\equal{\ALG@noend}{t}}{\algtext*{EndRequest}}{}
\makeatother
\algblockdefx[Upon]{Upon}{EndUpon}{\textbf{upon}~}{\textbf{end}}
\makeatletter
\ifthenelse{\equal{\ALG@noend}{t}}{\algtext*{EndUpon}}{}
\makeatother
\newcommand{\send}{\textbf{send to}~}
\algblockdefx[AsyncFun]{AsyncFun}{EndAsyncFun}{\textbf{async function}~}{\textbf{end}}

\graphicspath{ {figures/} }
\usepackage{tikz}
\usepackage[eulergreek]{sansmath}
\usepackage{pgfplots}
\usepackage{pgfplotstable}
\usepackage{cleveref}
\usepackage{textcomp}
\usepackage{comment}
\crefname{assumption}{assumption}{assumptions}

\pgfplotsset{compat=newest}
\usepgfplotslibrary{external,units,colorbrewer,groupplots,fillbetween,statistics}
\tikzsetexternalprefix{figures/}
\tikzset{external/mode=list and make}
\usetikzlibrary{patterns,shapes.misc}

\makeatletter
\begingroup\endlinechar=-1\relax
\everyeof{\noexpand}%
\edef\x{\endgroup\def\noexpand\homepath{%
		\@@input|"kpsewhich --var-value=HOME" }}\x
\makeatother

\def\overleafhome{/tmp}
\newcommand{\inputplot}[2]{%
	\ifx\homepath\overleafhome%
	\IfBeginWith{#1}{plots}{\includegraphics{main-figure#2.pdf}}{#1}%
	\else%
	{\sffamily\scriptsize\input{#1}}
	\fi
}

\newcommand{\newgroupwidth}[2]%
{\expandafter\xdef\csname groupwidth#1\endcsname{#2}}

\newcounter{groupwidth}
\newsavebox{\groupwidthbox}
\makeatletter
{\edef\groupnumber{#1}%
	\stepcounter{groupwidth}%
	\@ifundefined{groupwidth\thegroupwidth}{\pgfmathsetlengthmacro{\mywidth}{\linewidth/\groupnumber}}%
	{\expandafter\let\expandafter\mywidth\csname groupwidth\thegroupwidth\endcsname}%
	\begin{lrbox}{\groupwidthbox}%
		\tikzset{/pgfplots/width={\mywidth}}%
		\ignorespaces}%
	{\end{lrbox}%
	\usebox\groupwidthbox
	\pgfmathsetlengthmacro{\mywidth}{\mywidth + (\linewidth - \wd\groupwidthbox)/\groupnumber}
	\immediate\write\@auxout{\string\newgroupwidth{\thegroupwidth}{\mywidth}}}
\makeatother

\usepackage{amsmath,amssymb,amsfonts,amsthm}
\usepackage{physics}
\usepackage{enumitem}
\usepackage{thmtools} 
\usepackage{thm-restate}
\usepackage{bbm}
\theoremstyle{definition}

\theoremstyle{remark}

\allowdisplaybreaks

\copyrightyear{2025}
\acmYear{2025}
\setcopyright{acmlicensed}\acmConference[EuroMLSys '25]{The 5th Workshop on Machine Learning and Systems }{March 30-April 3 2025}{Rotterdam, Netherlands}
\acmBooktitle{The 5th Workshop on Machine Learning and Systems (EuroMLSys '25), March 30-April 3 2025, Rotterdam, Netherlands}
\acmDOI{10.1145/3721146.3721938}
\acmISBN{979-8-4007-1538-9/2025/03}


\begin{document}

\title{Practical Federated Learning without a Server}

\author{Akash Dhasade}
\affiliation{
  \institution{EPFL}
  \city{Lausanne}
  \country{Switzerland}
}

\author{Anne-Marie Kermarrec}
\affiliation{
  \institution{EPFL}
  \city{Lausanne}
  \country{Switzerland}
}

\author{Erick Lavoie}
\affiliation{
  \institution{University of Basel}
  \city{Basel}
  \country{Switzerland}
}

\author{Johan Pouwelse}
\affiliation{
  \institution{Delft University of Technology}
  \city{Delft}
  \country{The Netherlands}
}

\author{Rishi Sharma}
\affiliation{
  \institution{EPFL}
  \city{Lausanne}
  \country{Switzerland}
}

\author{Martijn de Vos}
\affiliation{
  \institution{EPFL}
  \city{Lausanne}
  \country{Switzerland}
}

\renewcommand{\shortauthors}{Dhasade et al.}

\begin{abstract}

Federated Learning (FL) enables end-user devices to collaboratively train ML models without sharing raw data, thereby preserving data privacy.
In FL, a central parameter server coordinates the learning process by iteratively aggregating the trained models received from clients. 
Yet, deploying a central server is not always feasible due to hardware unavailability, infrastructure constraints, or operational costs.
We present Plexus, a fully decentralized FL system for large networks that operates without the drawbacks originating from having a central server.
Plexus distributes the responsibilities of model aggregation and sampling among participating nodes while avoiding network-wide coordination.
We evaluate Plexus using realistic traces for compute speed, pairwise latency and network capacity.
Our experiments on three common learning tasks and with up to \num{1000} nodes empirically show that Plexus reduces time-to-accuracy by 1.4-1.6$\times$, communication volume by 15.8-292$\times$ and training resources needed for convergence by 30.5-77.9$\times$ compared to conventional decentralized learning algorithms.
\end{abstract}


\maketitle

\section{Introduction}
\label{sec:introduction}

\Ac{FL} enables devices (referred to as \emph{nodes} in this work) to collaboratively train a global \ac{ML} model without sharing their private training data. 
In a single \ac{FL} training round, a central server first selects a \textit{sample}, \ie, a random subset of online nodes, that train a model in parallel~\cite{pmlr-v54-mcmahan17a}.
Nodes then send their updated model to the server, which aggregates incoming models into a single model.
\ac{FL} is widely used today in various applications, including next-word prediction on keyboards~\cite{RN154,RN207,RN152}, speech recognition~\cite{RN209}, human activity recognition~\cite{RN160}, and healthcare~\cite{RN210, RN170, RN212, RN169}.

However, deploying and maintaining a central \ac{FL} server can be challenging for various reasons~\cite{yuan2024decentralized,sabuhi2024micro}.
Infrastructure constraints in remote or underdeveloped areas may make it difficult to set up a reliable central server.
Additionally, regulatory and privacy concerns might restrict centralized model aggregation, particularly in sensitive domains like healthcare and finance~\cite{nguyen2022federated}.
Furthermore, the reliance on a central server can introduce a single point of failure, stalling training progression if the server becomes unavailable~\cite{9700624,kang2020reliable,zhang2020federated}.
Despite these challenges, centralized model aggregation used by \ac{FL} offers an advantage over existing \ac{DL} approaches that rely on neighborhood-based aggregation since \ac{FL} results in higher model accuracy and faster model convergence.
We thus ask ourselves the question: \emph{Can we perform \ac{FL} without a server?}

We answer affirmatively and present \sys, a fully decentralized approach for \ac{FL} training.
The core of \sys lies in its \textit{decentralized peer sampler} that enables nodes to independently determine a subset of other nodes, or a \emph{sample}, that is in charge of the training process for a given round.
The sample changes every round, therefore evenly balancing the training load among nodes and providing nodes with an equal opportunity to contribute to model training.
Following local model training, nodes in a sample select a single \textit{aggregator} from the same sample that aggregates all trained models generated in each round.
This aggregator  then sends the aggregated model to nodes in the next sample, initiating the next round of training.

We evaluate \sys using real-world mobile phone traces of pairwise latencies, bandwidth capacities, and computation speeds~\cite{lai2022fedscale}.
Our evaluation covers three common learning tasks in varying network sizes, up to \num{1000} nodes.
We compare the performance of \sys against standard \ac{FL} and two baseline \ac{DL} algorithms: \dpsgd~\cite{lian2017can} and \ac{GL}~\cite{hegedHus2019gossip}.
Our experimental results show that \sys, compared to the best performing \ac{DL} baseline, reduces time-to-accuracy by 1.4-1.6$\times$, communication volume by 15.8-292$\times$, and training resources consumed by 30.5-77.9$\times$.
Furthermore, we demonstrate that \sys competes with the performance of \ac{FL} in real-world settings.

This work makes the following two contributions:

\begin{enumerate}
    \item We design \sys, a practical and decentralized \ac{FL} system (\Cref{section:system}).
    \sys incorporates a decentralized peer sampler to select a small subset of nodes that train the model each round, significantly reducing training resources required to converge compared to existing \ac{DL} approaches.
    Our system operates without any centralized or network-wide coordination.
    
    \item We implement \sys and conduct an extensive evaluation of \sys using real-world mobile phone traces at scale, comparing it with prominent baseline \ac{DL} algorithms and standard \ac{FL} with a server (\Cref{sec:eval}). Our results demonstrate that \sys, compared to \ac{DL} baselines, significantly enhances performance across three common learning tasks regarding time-to-accuracy, communication volume, and resource requirements, while providing similar accuracy to \ac{FL}.
\end{enumerate}

\begin{figure}[t]
    \centering
    \includegraphics[width=.65\linewidth]{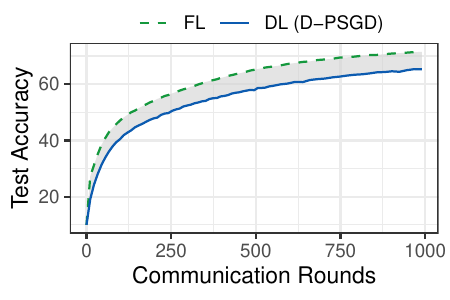}
    \caption{The evolution of test accuracy of \Ac{FL} and \ac{DL} (D-PSGD) on the \cifar dataset in a \num{1000}-node network. \Ac{FL} converges quicker and to higher accuracy than \DL.}
    \label{fig:motivation_aggregation}
\end{figure}

\section{Motivation}
\label{sec:motivation}

\textbf{\Acf{FL}} is a collaborative \ac{ML} algorithm where a central parameter server orchestrates the training of \ac{ML} models on client devices.
The effectiveness and practicality of \ac{FL} has been well-documented in various settings~\cite{RN154,RN209,RN160,RN210, RN170, RN212, RN169}.
However, its heavy reliance on a central server to coordinate training through \emph{client sampling} and sample-wide \emph{gradient aggregation} makes \ac{FL} impractical in many real-world scenarios.
In particular, \ac{FL} training can last for days, and the central server needs to remain continuously available throughout the training process.
Furthermore, the central parameter server in \ac{FL} demands a high-bandwidth network connection to communicate with multiple clients simultaneously.
These high availability and network bandwidth requirements result in significant operational and infrastructure costs for the \ac{FL} infrastructure.

\textbf{\DL and Residual Variance.} \DL emerges as a promising alternative to \ac{FL}~\cite{beltran2023decentralized}.
\Ac{DL} approaches like \Ac{D-PSGD}~\cite{lian2017can}, \Ac{EL}~\cite{de2024epidemic}, \Ac{GL}~\cite{ormandi2013gossip}, and derivatives~\cite{biswas2025noiseless,biswas2025boosting} eliminate the need for the central server by introducing peer-to-peer communication and model aggregation.
However, these \DL approaches often do not attain the same accuracies as \ac{FL}.
This is because nodes in \DL aggregate models amongst neighborhoods, \ie, local aggregation~\cite{lian2017can}.
Local aggregation leaves \emph{residual variance} between local models, which biases gradient computations and slows down model convergence compared to performing a global aggregation before starting a training round~\cite{bellet2021d,pmlr-v139-kong21a}.

To further illustrate this effect, we chart in~\Cref{fig:motivation_aggregation} the test accuracy for \FL and \dpsgd as the training progresses, on the \cifar dataset under an Independent and Identically Distributed (IID) data distribution in a network with \num{1000} nodes.
We implement \dpsgd with the state-of-the-art one-peer exponential graph topology (OP-Exp.)~\cite{ying2021exponential}.
With this topology, each node receives and sends exactly one model every round.
A peer is connected to $ log(n) $ neighbors ($ n $ is the total network size) and cycles through them round-robin.
All other learning parameters follow our experimental setup described in \Cref{sec:experiment_setup}.
\Cref{fig:motivation_aggregation} shows that \FL's aggregation is beneficial for convergence and reaches higher accuracy than \DL by nullifying residual variance across nodes in the network.

\section{Design of \sys} %
\label{section:system}

We first describe our system model and assumptions in Section~\ref{section:system-model}, provide a conceptual overview of the algorithm in Section~\ref{section:overview}, and then present the components of \sys in the remaining sections.%

\begin{figure*}[t]
    \centering
    \includegraphics[width=.7\linewidth]{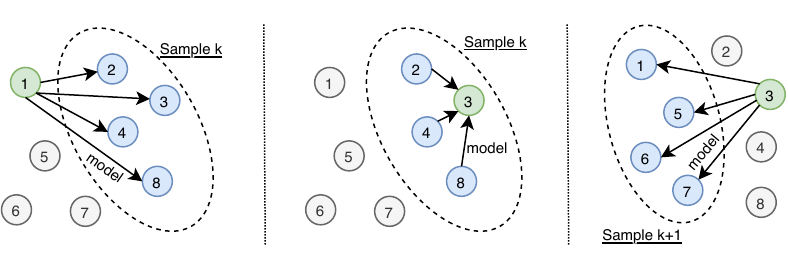}
    \caption{Overview of round $ k $ and $ k+1 $ in \sys, with 8 nodes and a sample size of 4 ($s = 4 $).
    Participants are indicated in blue and the aggregator in green.}
    \label{fig:midas_overview}
\end{figure*}

\subsection{System Model and Assumptions}
\label{section:system-model}

We consider a peer-to-peer network of $n$ nodes that collaboratively train a global \ac{ML} model $ \theta $.
Each participating node has access to a local dataset which never leaves the participants' device. 
Only the model parameters are exchanged between participating nodes. We assume that each node knows the specifications of the ML model being trained, the learning hyperparameters, and the settings specific to \sys.
These specifications should be exchanged before training starts.

This work focusses on \ac{FL} training in a cross-device setting without a central server.
Thus, model training with \sys proceeds in a decentralized environment and relies on the cooperation of nodes with varying resource capacities.
We assume that each node has a unique identifier (\eg, a public key) and assume that nodes are connected through a fully connected overlay network (\textit{i.e.}, all nodes can communicate with each other).
Though the computational capacities of participating nodes may vary, we assume that each node's computational resources are sufficient to reliably participate in learning.
We assume that the model being trained fits into the memory of each node.
Also, aggregators in \sys should have sufficient memory or disk space to store and aggregate the trained models produced by other nodes.
While we remark that nodes might act maliciously during the training process, we consciously leave out these considerations as this requires additional mechanisms.
However, we acknowledge research in privacy-preserving ML, many of which we believe could be integrated or adapted into \sys~\cite{bonawitz2017practical,mothukuri2021survey,biswas2025noiseless}.

\subsection{\sys in a Nutshell} %
\label{section:overview} 

Similar to \ac{FL}, \sys \textit{(i)} has a subset of nodes (a \emph{sample}) train the model each round, and \textit{(ii)} refreshes samples each round.
We refer to nodes belonging to a sample as \emph{participants}.
Among the participants, one node, named the \emph{aggregator}, is responsible for model aggregation during that round.
This aggregator is selected to be the node with the highest bandwidth capacity as it has to temporarily handle incoming model transfers from all participants in a round.
In each round, participants are randomly sampled from all nodes using a consistent hashing scheme.
This sampling mechanism is a key contribution of \sys and is further discussed in~\Cref{subsec:deriving_samples}.

\Cref{fig:midas_overview} illustrates two rounds (round $ k $ and $ k + 1 $) in \sys, with a network containing 8 nodes and a sample size of 4.
We denote the set of nodes in the $k$-th sample as $ S^k $ and the aggregator within $ S^k $ as $ a^k $. 
At the beginning of round $ k $, the aggregator in the previous sample $ S^{k-1} $ sends the aggregated model to all participants in $ S^k $ (step 1).
The participants train the model with their local data and then send their updated model to aggregator $a^{k}$ (step 2).
$a^{k}$ finally sends the aggregated model to the participants in sample $ S^{k+1} $, initiating round $ k + 1 $ (step 3).
This simplified algorithm description hinges on the ability of nodes to derive samples.
Thus, the main technical challenge lies in deriving consistent samples in a decentralized fashion.

\subsection{Deriving Samples and Aggregators}
\label{subsec:deriving_samples}

One of the main novelties of \sys is to decentralize the sampling procedure by having each participant in a sample compute the next sample independently. %
In order to achieve this, each node maintains a \emph{local view} of the network wherein the membership information of (all) other nodes is recorded.
The gist of \sys's sampling procedure is to rely on a hash function parameterized by the round number and the node identifiers, stored by all nodes in their local view so that each node can independently compute the sample of nodes expected to be active during the training. 
\Cref{alg:get_sample} shows the \sys sampling procedure, which aims to obtain a sample of $s$ currently active nodes in the $k^{th}$ round.
First, a subset of \textit{candidates} is retrieved. %
Concatenating the node identifiers with round numbers randomizes the order of nodes every round. 
The resulting list is sorted in lexicographic order, which provides the order in which candidates are contacted.
As long as the list of candidates is the same between all nodes, the order of contact and the resulting samples will mostly be the same. %

\begin{algorithm}[t]
\small
	\caption{Determining a sample and aggregator by node $i$ where $k$ denotes the round number and $s$ is the requested sample size.}
	\label{alg:get_sample}
	\begin{algorithmic}[1]

	    \Procedure{Sample}{$k,s$}
	        \State $H \leftarrow \textsc{sort}([ \textsc{hash}(j + k) ~\textbf{for}~ j ~\textbf{in}~ \textsc{Nodes}() ])$
	        
    	    \State $C \leftarrow [ j~\textbf{for}~h_j~\textbf{in}~H ]$
                \State \textbf{return} $C[:s]$
    	    
	    \EndProcedure
        \State
        \Procedure{Aggregator}{$k,s$}
		  \State $S^k \leftarrow \textsc{Sample}(k,s)$
		  \State \Return $j \in S^k $ \textbf{such that} $j$ has the largest bandwidth among all $S^k$ nodes according to $B_i$
		\EndProcedure
        
	\end{algorithmic}
\end{algorithm}

\begin{algorithm}[ht]
    \small
	\caption{Training and aggregation by node $i$.}
	\label{alg:training}
	\begin{algorithmic}[1]
        \State \textbf{Require}: Sample size $s$, success fraction $sf$
        \State $\Theta \leftarrow []$ \Comment{List of received models as aggregator}
        \State
		\If{$i~\textbf{in}~\textsc{sample}(1, s)$} \Comment{Start training if we are in the first sample}
		\State \send $i$ \texttt{train}$(1, \textsc{randomModel}())$
		\EndIf
		
		\State
		\Upon \texttt{train}$(k, \theta)$ \Comment{Training}
		\State $\bar \theta \leftarrow \textsc{train}(\theta)$
        \State $a \leftarrow \textsc{Aggregator}(k, s)$ \Comment{Alg. \ref{alg:get_sample}}
		\State \send $a$ \texttt{aggregate}$(k, \bar \theta)$
		\EndUpon

        \State
        \Upon \texttt{aggregate}$(k,\theta_j)$ \textbf{from} $ j $ \Comment{Aggregation}
        \State $\Theta.\textsc{add}(\theta_j)$
        \If{$\Theta.\textsc{size} \geq \lfloor s \times sf \rfloor $ }
        \For{\textbf{all} $j$~\textbf{in}~ $S^{k+1}$ ~\textbf{in parallel} }
        \State $ \theta_{agg} \leftarrow~\textsc{avg}(\Theta) $
		\State \send $j$ \texttt{train}$(k+1, \theta_{agg} )$
		\EndFor
		\State $\Theta \leftarrow []$ \Comment{Reset list}
        \EndIf
        \EndUpon
        
	\end{algorithmic}
\end{algorithm}

We next discuss how participants determine an aggregator.
Internally, this method calls the sampling procedure from \Cref{alg:get_sample}.
The aggregator is a critical node for system progression and must handle the reception and transmission of at most $ s $ trained models during a round.
Since the aggregator has to handle this network load, we preferentially select the participant with the highest bandwidth capacity from the derived sample, but acknowledge that other selection strategies are possible.
This biased aggregator selection optimizes the model transfer times and reduces the time required per round compared to uniform aggregator selection.
We remark that this biased selection has no impact on the quality of the trained model.
Bandwidth capacities of individual nodes can be determined and synchronized a-priori to training.
We found that this decision was essential to the success of \sys as learning progress would slow down significantly if an aggregator with low bandwidth capacity is chosen, especially if the model size increases.

\subsection{Training and Aggregating Models}
\label{subsection:training-aggregating}
Each node in \sys implements two procedures: one for aggregation and one for training.
We provide the pseudocode of these procedures in \Cref{alg:training}.
The design of \sys is based on a \textit{push}-based architecture, in which nodes in sample $S^k$ trigger the activation of nodes in sample $S^{k+1}$. 
This way, nodes do not have to continuously be aware of the current training round being worked on; they only have to start working when receiving a trained or aggregated model. 

\textbf{Training.} We first describe the training procedure by node $ i $.
A node starts the training task when it receives a \texttt{train} message containing an aggregated model $ \theta $ and a round number $ k $.
$ i $ trains the received model $ \theta $ by calling the \textsc{train} procedure, resulting in trained model $ \bar \theta $.
Then, $ i $ determines aggregator $ a $ in current sample $ k $ by calling the \textsc{aggregate} procedure.
This aggregator is derived using our peer sampler (see \Cref{subsec:deriving_samples} and \Cref{alg:get_sample}).
Finally, $ i $ sends an \texttt{Aggregate} message, containing the resulting model $ \bar \theta $, to aggregator $ a $.

\textbf{Aggregation.}
We provide the pseudocode related to aggregation in \Cref{alg:training}.
An aggregator $ a $ starts the aggregation task when it is activated through an \texttt{aggregate} message from node $ j $, containing trained model $ \theta_j $.
$ a $ keeps track of the received models in list $ \Theta $ and adds $ \theta_j $ to $\Theta $.
Upon receiving at least $ \lfloor s \times sf \rfloor $ models for round $ k $, $ a $ aggregates these models, resulting in $\theta_{agg} $.
We refer to the required fraction of models needed as the \emph{success fraction} $ sf $.
This oversampling is common in realistic \ac{FL} systems~\cite{abdelmoniem2023refl}.
$ a $ then determines the participants in sample $ k+1 $ and sends these nodes a \texttt{train} message containing the next round number $ k + 1$ and the aggregated model $ \theta_{agg} $.
This completes round $ k $.

\subsection{\sys and \ac{FL}}

\sys brings \ac{FL} to fully decentralized networks.
We now discuss how the elements of \sys translate to those of \ac{FL}.
Firstly, the local model training at the nodes in \sys is equivalent to that in \ac{FL}. %
The aggregator in \sys temporarily performs the role of the \ac{FL} server, aggregating the model updates in the current round. %
The central server in \ac{FL} also orchestrates training by choosing the participants in each round.
This role is handled by the decentralized sampling mechanism in \sys described in \Cref{subsec:deriving_samples}.
As a consequence of the similarity of \sys to \ac{FL}, the convergence proofs for \ac{FL} also offer theoretical grounding for the convergence of our approach~\cite{li2019convergence}.

\begin{figure*}[t]
	\centering
     \begin{subfigure}{.55\textwidth}
    	\centering
    	\includegraphics[width=\linewidth]{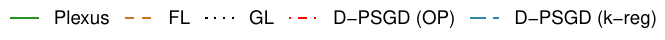}
    \end{subfigure}
	\begin{subfigure}{.33\linewidth}
		\centering
		\includegraphics[width=\linewidth]{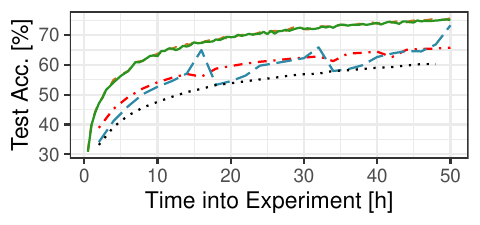}
		\label{fig:exp_accuracy_cifar10_iid}
	\end{subfigure}%
	\begin{subfigure}{.33\linewidth}
		\centering
		\includegraphics[width=\linewidth]{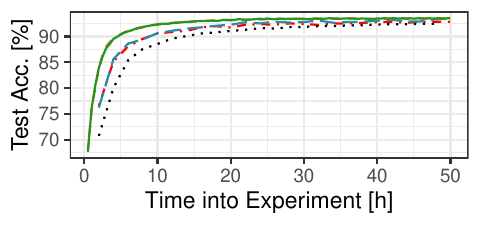}
		\label{fig:exp_accuracy_celeba}
	\end{subfigure}%
	\begin{subfigure}{.33\linewidth}
		\centering
		\includegraphics[width=\linewidth]{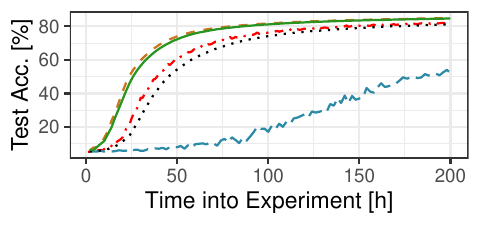}
		\label{fig:exp_accuracy_femnist}
	\end{subfigure}\vspace{-15pt}
	\begin{subfigure}{.33\linewidth}
		\centering
		\includegraphics[width=\linewidth]{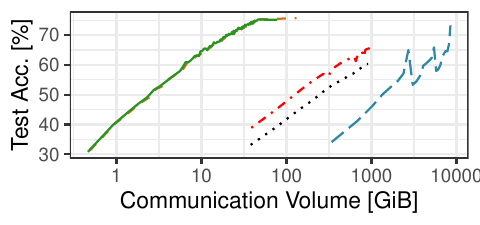}
		\label{fig:exp_communication_cifar10_iid}
	\end{subfigure}%
	\begin{subfigure}{.33\linewidth}
		\centering
		\includegraphics[width=\linewidth]{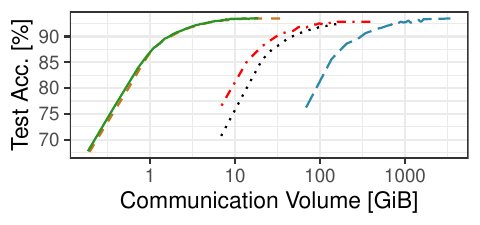}
		\label{fig:exp_communication_celeba}
	\end{subfigure}
	\begin{subfigure}{.33\linewidth}
		\centering
		\includegraphics[width=\linewidth]{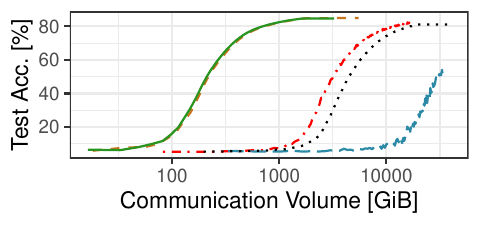}
		\label{fig:exp_communication_femnist}
	\end{subfigure}\vspace{-15pt}
	\begin{subfigure}{.33\linewidth}
		\centering
		\includegraphics[width=\linewidth]{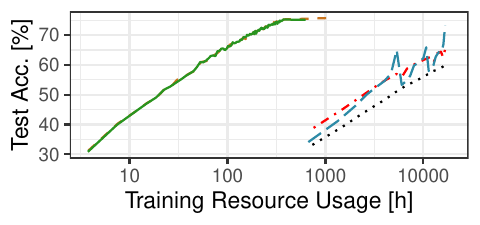}
		\caption{\cifar (IID)}
		\label{fig:exp_resource_cifar10_iid}
	\end{subfigure}%
	\begin{subfigure}{.33\linewidth}
		\centering
		\includegraphics[width=\linewidth]{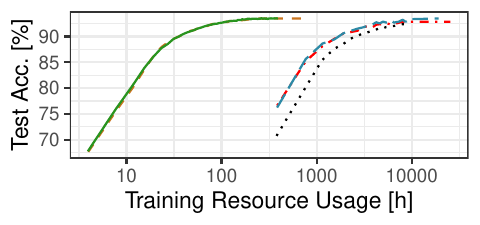}
		\caption{\celeba (non-IID)}
		\label{fig:exp_resource_celeba}
	\end{subfigure}%
	\begin{subfigure}{.33\linewidth}
		\centering
		\includegraphics[width=\linewidth]{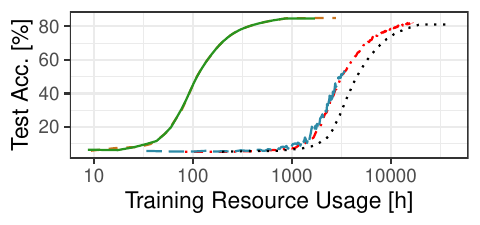}
		\caption{\femnist (non-IID)}
		\label{fig:exp_resource_femnist}
	\end{subfigure}
	\caption{The convergence of \sys and baselines against time (top), communication volume (middle), and training resource usage (bottom).}
	\label{fig:exp_comparison_fl}
\end{figure*}

\section{Experimental Evaluation}
\label{sec:eval}
We now present the experimental evaluation of \sys.
We provide all relevant details on our experiment setup in~\Cref{sec:experiment_setup}.
Our evaluation answers the following two questions: 
\begin{itemize}
    \item What is the performance of \sys in terms of wall-clock convergence time, communication volume and training resource usage compared to \ac{FL} and \ac{DL} baselines (\Cref{sec:exp_comparison_fl})?
    \item What is the effect of the sample size $ s $ in \sys on wall-clock convergence time, communication volume and training resource usage (\Cref{sec:exp_vary_s})?
\end{itemize}

\subsection{Experiment Setup}
\label{sec:experiment_setup}
We implement \sys in the Python 3 programming language.\footnote{Source code: \url{https://github.com/sacs-epfl/plexus}.}
\sys leverages the \texttt{IPv8} networking library which provides support for authenticated messaging and building decentralized overlay networks~\cite{ipv8}.
Our implementation adopts an event-driven programming model with the \texttt{asyncio} library.
We use the \texttt{PyTorch} library~\cite{paszke2019pytorch} to train ML models, and the dataset API from \textsc{DecentralizePy}~\cite{dhasade2023decentralized}.
As a node might be involved in multiple incoming and outgoing model transfers simultaneously, we equip each node with a bandwidth scheduler that we implemented.
This scheduler coordinates all model transfers a particular node is involved in and enables us to realistically emulate the duration of model transfers.

\textbf{Hardware.}
We run all experiments on machines in our national compute cluster.
Each machine is equipped with dual 24-core AMD EPYC-2 CPU, 128 GB of memory, an NVIDIA RTX A4000 GPU, and is running CentOS 8.
Similar to related work in the domain, we simulate the passing of time in our experiments~\cite{lai2022fedscale,lai2021oort,abdelmoniem2023refl}.
We achieve this by customizing the default event loop provided by the \texttt{asyncio} library and processing events without delay.
The real-world time required to reproduce our experiment depends on the baseline being evaluated.
Running our \sys and our \ac{FL} baseline requires between 6 hours (for \celeba) and 24 hours (for \femnist) of compute time.
The \ac{DL} baselines are more computationally intense since they have every node training each round and require between 24 hours (for \celeba) and 5 days (for \femnist) of compute time.
Our simulator is implemented as a single process and its efficiency can be further improved by paralellizing the training.

\begin{table}[t!]
	\small
    \caption{Summary of datasets and learning parameters used to evaluate \sys and \ac{DL} baselines. ``Mom.'' denotes the momentum parameter.}
	\centering
	\begin{tabular}{c|c|c|c}
		\toprule
		\textsc{Dataset} & \textsc{Nodes} & \textsc{Learning Parameters} & \textsc{Model}  \\ \hline
		\cifar~\cite{krizhevsky2009learning} & \num{1000} & $ \eta = 0.002 $, mom. $ = 0.9 $ & CNN~\cite{hsieh2020non} \\ 
		\celeba~\cite{caldas2018leaf} & 500 & $ \eta = 0.001 $ & CNN\\ 
		\femnist~\cite{caldas2018leaf} & 355 & $ \eta = 0.004 $ & CNN\\ 
		\bottomrule
	\end{tabular}
	\label{table:experiment_datasets}
\end{table}

\textbf{Traces.}
We have designed \sys to operate in highly heterogeneous environments, such as mobile networks.
To verify that \sys also functions in such environments, we adopt various real-world traces to simulate pairwise network latency, bandwidth capacities, compute speed, and availability.
To model a WAN environment, we apply latency to outgoing network traffic at the application layer to realistically model delays in sending \sys control messages.
To this end, we collect ping times from WonderNetwork, providing estimations on the RTT between their servers located in 227 geo-separated cities~\cite{WonderNetwork}.
When starting an experiment, we assign peers to each city in a round-robin fashion and delay outgoing network traffic accordingly.

We also adopt traces from the \textsc{FedScale} benchmark to simulate the hardware performance of nodes, specifically network and compute capacities~\cite{lai2022fedscale}.
These traces contain the hardware profile of \num{131000} mobile devices and are originally sourced from the AI benchmark~\cite{ignatov2019ai} and the \textsc{MobiPerf} measurements~\cite{huang2011mobiperf}.
We assume that nodes are aware of the bandwidth capabilities of other nodes, and within a sample, a node sends its trained model to the aggregator with the highest bandwidth capability in the next sample.
In summary, our experiments go beyond existing work on \ac{DL} by integrating multiple traces that together account for the system heterogeneity in WAN environments.

\textbf{Datasets.}
We evaluate \sys on different models and with three distinct datasets, whose characteristics are displayed in Table~\ref{table:experiment_datasets}.
The \cifar dataset~\cite{krizhevsky2009learning} is IID, partitioned by uniformly randomly assigning data samples to nodes. %
The \celeba and \femnist datasets are taken from the \leaf benchmark~\cite{caldas2018leaf}, which was specifically designed to evaluate the performance of learning tasks in non-IID settings.
The sample-to-node assignment for \celeba and \femnist is given by the \leaf benchmark.
Our evaluation, thus, covers a variety of learning tasks and data partitions.

\textbf{Performance Metrics and Hyperparameters.}
We measure the top-1 test accuracy of the model on a global test set unseen during training, for the purpose of evaluation.
In line with other work, we fix the batch size to 20 for all experiments and each device always performs five local steps when training its model~\cite{lai2022fedscale,abdelmoniem2023refl} before communicating.
All models are trained using the SGD optimizer.
For \cifar we additionally use a momentum factor of $0.9$.
All our learning parameters were adopted from previous works~\cite{caldas2018leaf, bellet2021d} or were considered after trials on several values.
They yield acceptable target accuracy for all evaluated datasets.
We run each experiment three times with different seeds and report averaged values.

\textbf{Baselines.}
We compare \sys against a \ac{FL} setup in which we assume the availability of a server with unlimited bandwidth constraints.
We also use \Acf{GL}~\cite{hegedHus2019gossip} and \dpsgd~\cite{lian2017can} as \ac{DL} baselines.
In each round of \ac{GL}, a node first waits for some time and then sends its model to another random node in the network.
The selection of nodes is facilitated by a peer-sampling service which presents a view of random nodes in the network every round.
Upon receiving a model from another node, the recipient node merges it with its own local model, weighted by the model age, and trains the local model. 
\ac{GL} naturally tolerates churn and is robust to failing nodes.
However, pairwise model aggregation still leaves residual variance and deteriorates model convergence compared to when using global aggregation.
In our experiments, we fix the round timeout to \num{60} seconds for \ac{GL} to give each node sufficient time to train and transfer the model each local round.

\dpsgd~\cite{lian2017can} is a synchronous algorithm that only proceeds when all nodes have received all models from their neighboring nodes.
We evaluate \dpsgd under two topologies: a 10-regular topology (\ie each node has ten neighbors) and a one-peer exponential graph topology, the latter being a state-of-the-art topology in \ac{DL}~\cite{ying2021exponential}.
Thus, we evaluate \dpsgd with sparse and dense graph connectivity.

For \sys, we report the accuracy of the global model after aggregation every ten rounds.
For \dpsgd, we determine the mean and standard deviation of the accuracy obtained by evaluating models of individual nodes on the test dataset every two hours. 
We also report communication volume (transmitted bytes) and training resource usage (\ie, the time a device spends on model training).
For \sys, we set $ s = 13 $ and adjust the aggregator so it proceeds when it has received 80\% of all models ($sf = 0.8$).
We run experiments with \cifar and \celeba for 50 hours and \femnist for 200 hours which gives model training with \sys and other baselines sufficient time to converge.

\subsection{\sys Compared to Baselines}
\label{sec:exp_comparison_fl}
We quantify and compare the performance of \sys with baseline systems.
We compare \sys against a \ac{FL} setup in which we assume the availability of \emph{a server with unlimited bandwidth constraints}.
We also use \Acf{GL}~\cite{hegedHus2019gossip} and \dpsgd~\cite{lian2017can} as \ac{DL} baselines.
We evaluate \dpsgd under two topologies: a 10-regular topology (\ie each node has ten neighbors) and a one-peer exponential graph topology, the latter being a state-of-the-art topology in \ac{DL}~\cite{ying2021exponential}.
Thus, we evaluate \dpsgd with both sparse and dense graph connectivity.

\begin{figure*}[t]
	\centering
     \begin{subfigure}{.3\textwidth}
    	\centering
    	\includegraphics[width=\linewidth]{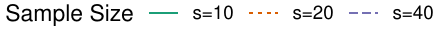}
    \end{subfigure}
	\begin{subfigure}{.33\linewidth}
		\centering
		\includegraphics[width=\linewidth]{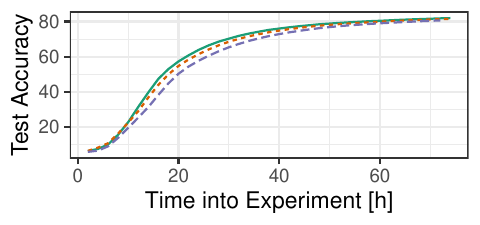}
        \caption{Model convergence}
		\label{fig:exp_vary_s_accuracy}
	\end{subfigure}%
	\begin{subfigure}{.33\linewidth}
		\centering
		\includegraphics[width=\linewidth]{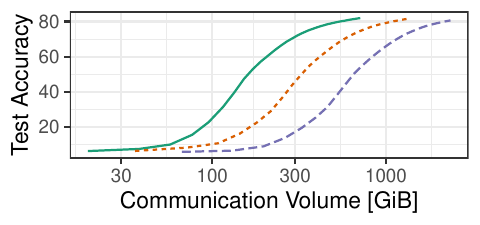}
        \caption{Communication Volume}
		\label{fig:exp_vary_s_communication}
	\end{subfigure}%
	\begin{subfigure}{.33\linewidth}
		\centering
		\includegraphics[width=\linewidth]{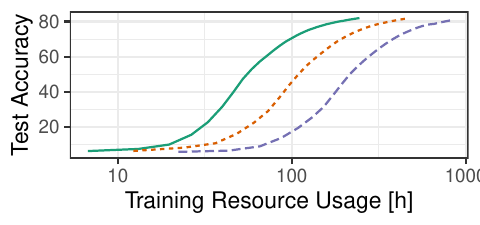}
        \caption{Training Resource Usage}
		\label{fig:exp_vary_s_resources}
	\end{subfigure}
	\caption{The performance of \sys for different sample sizes $s$, using the \femnist dataset.}
	\label{fig:exp_vary_s}
\end{figure*}

\textbf{Results.}
We show the performance of \sys and baselines in~\Cref{fig:exp_comparison_fl}.
The top row of~\Cref{fig:exp_comparison_fl} shows the test accuracy as the experiment progresses.
\sys outperforms both \ac{DL} baselines by converging quicker and achieving higher test accuracy, consistently across all datasets.
In general, we find that in \ac{GL} more training occurs within a given time unit compared to \ac{DL}, since \ac{GL} rounds are asynchronous and individual nodes have less idle time compared to \dpsgd.
On the simpler binary classification task for the \celeba dataset, the performance improvements of \sys are modest.
However, on more difficult learning tasks like the 62-class image classification in \femnist with a larger model size, \sys achieves more than 20\% better accuracy when compared to the best performing \ac{DL} baseline, \ac{GL}.
We also observe that \sys generally shows comparable time-to-accuracy as \ac{FL} but with the larger model size of \femnist we observe small differences since the server in \ac{FL} has unlimited bandwidth.
The middle row in~\Cref{fig:exp_comparison_fl} shows the communication volume (horizontal axis in log scale) required to achieve the test accuracy for the evaluated systems.
\sys attains high test accuracies with orders of magnitude less transmitted bytes.
These efficiency gains are particularly pronounced for the \femnist dataset.
We note that \dpsgd with a 10-regular topology incurs the most network traffic while performing on par or worse than the one-peer exponential graph topology.
The bottom row in~\Cref{fig:exp_comparison_fl} shows the training resource usage (horizontal axis in log scale) consumed to achieve the test accuracy.
\sys attains high test accuracies with orders of magnitude less resource usage.
Finally, we note that \sys incurs comparable communication volume and resource usage as the centralized baseline of \ac{FL}.

For each dataset, we determine the best-performing baseline in terms of the highest \emph{individual model accuracy achieved across all nodes}.
We then determine time-to-accuracy (TTA), communication-to-accuracy (CTA), and resources-to-accuracy (RTA), which are metrics that represent the efficacy, efficiency, and scalability of \ac{DL} systems.
For the evaluated datasets and compared to the target accuracy, \emph{\sys saves 1.4$\times$ - 1.6$\times$ in TTA, 15.8$\times$ - 292$\times$ in CTA and 30.5$\times$ - 77.9$\times$ in RTA compared to \ac{GL} and \dpsgd}.
In conclusion, our comprehensive evaluation demonstrates the superior efficiency and effectiveness of \sys. %

\begin{figure}[t]
    \centering
    \includegraphics[width=.75\linewidth]{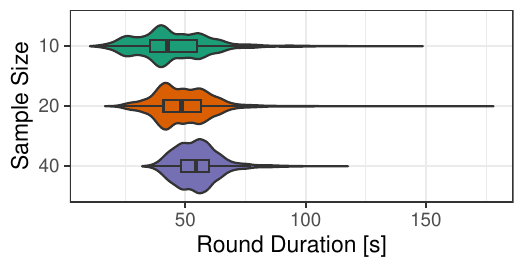}
    \caption{The distribution of round durations, for the \femnist dataset and different sample sizes.}
    \label{fig:vary_s_round_durations}
\end{figure}

\subsection{Varying the Sample Size $s$}
\label{sec:exp_vary_s}
We next explore the effect of the sample size $ s $ on model convergence, communication volume, training resource usage, and round duration by running \sys with $s = 10$, $20$, and $40$. 
This experiment uses the \femnist dataset which has the largest model size in our setup.
\Cref{fig:exp_vary_s_accuracy} shows the test accuracy for the three different values of $ s $ as the experiment progresses.
We observe that increasing $ s $ actually slows down training, likely because more models have to be transferred to and from the aggregator.
Naturally, increasing $ s $ also has a negative impact on communication cost and resource usage, which are visualized in~\Cref{fig:exp_vary_s_communication} and~\Cref{fig:exp_vary_s_resources}, respectively.
In particular, reaching 80\% test accuracy for $ s = 40 $ incurs 4.0$\times$ additional communication volume and 4.0$\times$ more training resource usage, when compared to $ s = 10 $.
We note, however, that increasing $ s $ might be favorable in other scenarios, \eg when data is highly non-IID.

Increasing $ s $ also prolongs the duration of individual rounds as the aggregator has to receive and redistribute more models.
We show in~\Cref{fig:vary_s_round_durations} the distribution of round durations in seconds for different values of $ s $ using a box and violin plot.
When increasing $ s $ from 10 to 40, the average round duration increases from \SI{45.1}{seconds}. to \SI{54.9}{seconds}.
Moreover, we observe that some rounds take disproportionally long which can happen when all nodes in a selected sample have low bandwidth or slow compute speeds.
For example, a round for $ s = 20 $ took up up to \SI{178}{seconds}.
However, this is relatively rare: for $ s = 20 $, only 18 rounds out of \num{6941} took over \SI{100}{seconds} to complete.
We observe also a positive effect on round duration when increasing $ s $: with higher values of $ s $, there is a higher probability that nodes with high bandwidth capacities are included in the sample compared to lower values of $ s $, which lowers the overall model transfer times during a round.
We can see this effect in~\Cref{fig:vary_s_round_durations} as there is a higher variance in round durations for lower values of $ s $.
Empirically, we obtain a good trade-off between sample size and convergence when setting the sample size around 10.

\section{Related Work}
\label{sec:related}

\textbf{Decentralized Learning (DL).}
\dpsgd~\cite{lian2017can} showed theoretically and empirically that under strong bandwidth limitations on an aggregation server in a data center, decentralized algorithms can converge faster. %
Assumptions on the behavior of those algorithms make them most suited to data centers.
The synchronization required in \dpsgd is costly when training on edge devices.
As a solution, research in \ac{DL} has been focussed either on having a better topology~\cite{bellet2021d, de2024epidemic, song2022communicationefficient, takezawa2024beyond, vogels2022beyond, ying2021exponential}, or designing asynchronous algorithms~\cite{biswas2025boosting, pmlr-v80-lian18a, nabli2024textbf, ormandi2013gossip}.
MoshpitSGD~\cite{ryabinin2021moshpit} uses a DHT to randomly combine nodes in multiple disjoint cohorts every round for fast-averaging convergence.
Notably, Teleportation is a \ac{DL} algorithm where, similar to \sys, a small subset of nodes train every round and then exchange models with other sampled nodes over a smaller topology~\cite{takezawa2025scalable}.
However, the paper does not specify exactly how these nodes are sampled and \sys solves this issue through consistent hashing.
All the above algorithms, however, overlook system heterogeneity whereas we evaluated \sys in the presence of network and compute heterogeneity.

\textbf{Federated Learning (FL).}
\ac{FL} is arguably the most popular algorithm for privacy-preserving distributed learning and uses a parameter server to coordinate the learning process~\cite{pmlr-v54-mcmahan17a}.
Similar to \sys, FL lowers resource and communication costs at the edge by having a small subset of nodes train the model every round.
To make FL suitable at scale in deployment scenarios, recent works have placed significant emphasis on system challenges~\cite{zhang2023no,abdelmoniem2023refl,bonawitz2019towards,MLSYS2022_f340f1b1,lai2021oort,pmlr-v151-nguyen22b}.
FL still requires a highly available central server that can support many clients concurrently, possibly resulting in high infrastructure costs.
\sys, on the contrary, is a fully decentralized system with an aggregation scheme inspired by \ac{FL}, while avoiding central coordination.

\textbf{Blockchain-Assisted \ac{DL}.}
We identified various works that implement and evaluate blockchain-based decentralized learning systems~\cite{10330055,9274451,bao2019flchain,lu2019blockchain,majeed2019flchain,pokhrel2020federated} and discuss the challenges therein~\cite{zhang2024decentralized,TANG20242451}.
Consensus-based replicated ledgers used in these systems provide strong consensus primitives at the cost of a significant and unnecessary overhead.
Machine learning optimizations based on SGD thrive in the presence of stochasticity, obviating the need for strong consensus in the form of a global model~\cite{lian2017d-psgd,ormandi2013gossip}.

\textbf{Decentralized Peer Sampling.} Brahms~\cite{bortnikov2009brahms}, Basalt~\cite{auvolat2021scriptstyle}, PeerSampling~\cite{jelasity2007gossip} and PeerSwap~\cite{guerraoui2024peerswap} provide each node with a different, uniformly random sample without network-wide synchronization.
In contrast, the peer sampling of \sys instead ensures that nodes select \emph{equivalent} samples in a particular round of training.

\section{Broader Impact and Open Challenges}
\label{sec:broader_impact_and_limitations}
The broader impact of our work is multifaceted, addressing both a  technical and socio-economic dimension.
By circumventing the need for a central server, \sys can lower the barrier and costs to adopt \ac{FL}-like training in decentralized settings.
A complementary benefit of \sys is that it enhances data privacy and security by decentralizing the model aggregation process, reducing the risk of data breaches associated with centralized systems.
This is because, depending on the total network size and sample size, it becomes more difficult for a single node to systematically access model updates from particular nodes, raising the barrier for privacy attacks such as the gradient inversion attack~\cite{huang2021evaluating}.

\textbf{Open Challenges.}
We discuss several open challenges in the current design of \sys.
Like any decentralized system, \sys must balance efficiency with network constraints.
As model sizes increase, aggregators may experience heavier loads, potentially prolonging round durations.
This also applies to \ac{FL} with a server, although it is typical that the server coordinating the \ac{FL} process has more compute capacity than the devices participating in the training process.
To account for this bottleneck in \sys, we select the aggregator as the node with the highest bandwidth capabilities within a sample.
However, in rare cases, all nodes in a sample may have low bandwidth capabilities due to an unfortunate selection (also see \Cref{sec:exp_vary_s}).
In this scenario, we could select an aggregator from outside the current sample.
Another approach to reduce communication burdens on aggregators is to use multiple aggregators in the same round.
Both enhancements, however, require us to deviate from the standard \ac{FL} algorithm and necessitate additional coordination mechanisms.

Secondly, we currently assume all nodes remain online during training which is typically not the case in real-world settings.
Our idea is to extend \sys with support for churn by including the current online or offline status in the local views, and synchronize these views between samples.
However, dealing with churn requires additional coordination as nodes can also go offline during training or aggregation.

Thirdly, \sys overutilizes the computational resources.
In our experiments, the round completes when 80\% of the models are received, thus at least 20\% of trained models will never be aggregated.
As a result, some participant updates may not be aggregated in every round, similar to common \ac{FL} techniques that mitigate stragglers~\cite{bonawitz2019towards}.
Various \ac{FL} systems alleviate this issue by integrating stale model updates~\cite{wu2020safa,damaskinos2022fleet,abdelmoniem2023refl}.
We leave integrating this in \sys for future work.

Finally, securing \sys against Byzantine nodes remains an open challenge.
A promising direction is to incorporate accountability mechanisms where nodes verify the integrity of sample selection, ensuring honest participation.

Overall, these aspects highlight common challenges in \ac{FL} and \ac{DL}.
Addressing them will improve the robustness and reliability of \sys, paving the way for deployment of decentralized learning systems in open and large-scale settings.

\section{Conclusions}
\label{sec:conclusion}
This paper introduced \sys, a practical, efficient and decentralized \ac{FL} system.
The two key components of our system are \textit{(i)} a decentralized peer sampler to select small subsets of nodes each round, and \textit{(ii)} a global aggregation operation within these selected subsets.
Extensive evaluations with realistic traces of compute speed, network capacity, and availability in decentralized networks up to \num{1000} nodes demonstrate the superiority of \sys over baseline \ac{DL} algorithms, reducing time-to-accuracy by 1.4-1.6$\times$, communication volume by 15.8-292$\times$, and training resource usage by 30.5-77.9$\times$ compared to \ac{DL} baselines.
Moreover, \sys also achieves accuracy and resource usage comparable to a centralized \ac{FL} baseline.

\begin{acks}
This work has been funded by the Swiss National Science Foundation, under the project ``FRIDAY: Frugal, Privacy-Aware and Practical Decentralized Learning'', SNSF proposal No. 10.001.796.
This work has also been funded by the Dutch national NWO/TKI science grant BLOCK.2019.004.
\end{acks}

\bibliographystyle{plain}
\bibliography{main.bib}

\end{document}